# *Quantized Angular Momentum in Topological Optical Systems*

*Mário G. Silveirinha*[*]

[1] *University of Lisbon–Instituto Superior Técnico and Instituto de Telecomunicações, Avenida Rovisco Pais, 1, 1049-001 Lisboa, Portugal, mario.silveirinha@co.it.pt*

**Abstract**

The Chern index characterizes the topological phases of nonreciprocal photonic systems. Unlike in electronic systems, the photonic Chern number has no clear physical meaning, except that it determines the net number of unidirectional edge states supported by an interface with a trivial mirror. Here, we fill in this gap by demonstrating that the photonic Chern number can be understood as a *quantum* of the light-angular momentum in a photonic insulator cavity. It is proven that for a large cavity, when the discrete spectrum can be approximated by a continuum, the spectral density of the thermal fluctuation-induced angular momentum is precisely quantized in the band-gaps of the bulk states. The nontrivial expectation of the light angular momentum is due to a circulation of thermal energy in closed orbits. Remarkably, this result can be extended to systems without a topological classification, and in such a case the "quantum" of the angular momentum density is determined by the net number of unidirectional edge states supported by the cavity walls.

[*] To whom correspondence should be addressed: E-mail: *mario.silveirinha@co.it.pt*

# Main Text

"Topological matter" and topological effects have elicited a great deal of interest in recent years, first in electronics [1-5] and then in photonics and acoustics [6-13]. There are several sub-classes of photonic materials with topological properties [9-11], but only optical systems with a broken-time reversal symmetry provide a strong-topological protection [14, 15]. In such systems, the different topological phases are characterized by a topological index known as the Chern number.

In the fermionic case the Chern number has an immediate physical meaning: it determines the value of the quantized Hall conductivity of a 2D electron gas [4, 16], and thus the electronic transport properties in the zero temperature limit. In contrast, in photonics the Chern number has not been linked with any specific physical quantity, except that similar to electronics it determines the number of gapless unidirectional edge states at an interface with a trivial insulator [6, 8, 14, 15]. In this article, we fill in this gap by proving that the (thermal or quantum) fluctuation-induced light *angular momentum* in a generic photonic-insulator cavity in thermodynamic equilibrium with a large reservoir has a spectral density per unit of area that is quantized in units of $\frac{1}{\pi c^2}\mathcal{E}_{T,\omega}$, with $\mathcal{E}_{T,\omega} = \frac{\hbar\omega}{2}\coth\left(\frac{\hbar\omega}{2k_B T}\right)$ the mean energy of a harmonic oscillator at temperature $T$ [17]. The nontrivial light angular momentum is due to a circulation of "heat" in closed orbits [18]. This result is rather general and remarkably applies even to systems *without* any topological classification. In the case of topological systems, the quantized angular momentum spectral density is precisely determined by the photonic Chern number.

We consider a generic closed cavity filled with either a photonic crystal or, alternatively, an electromagnetic continuum with no intrinsic periodicity (Fig. 1). The cavity cross-section shape in the *xoy*-plane can be rather arbitrary, but for simplicity first we focus in a rectangular cross-section with dimensions $L_1 \times L_2$. The cavity's height along the *z*-direction is *d*. The cavity walls are assumed "opaque", i.e., impenetrable by light. For example, perfectly electric conducting (PEC) walls are "opaque". The cavity may be regarded as a parallel-plate waveguide (with propagation plane parallel to *xoy*) terminated with the lateral walls. The equivalent unbounded "waveguide" does not support electromagnetic states in the spectral range of interest ("photonic insulator"). Note that the band structure is determined not only by the materials inside the cavity but also by the top and bottom walls. For simplicity, the effects of material loss in the wave propagation are assumed negligible. Recently, it was shown that topological concepts may be extended to some non-Hermitian systems, e.g., systems with loss and or gain [19-22].

Importantly, even when there are no bulk states, the opaque boundaries may enable the propagation of edge states localized at the cavity lateral walls [18, 23]. Such edge states famously occur in topological platforms [6, 7, 9, 14, 15], but they may as well emerge in systems with no topological classification [24]. Furthermore, the edge-states may be supported even when the system is time-reversal invariant (reciprocal) [9, 23, 24].

Consider a generic edge-state circulating around the cavity lateral walls (Fig. 1). The circulating motion is evidently associated with a non-trivial light angular momentum given by $\mathcal{L} = \frac{1}{c^2} \int dV\, \mathbf{r} \times \mathbf{S}$, where $\mathbf{S}$ is the Poynting vector, $\mathbf{r} = (x, y, z)$ is a generic point in space, and the integration is over the cavity volume. The angular momentum of photonic systems is extensively discussed in Refs. [25-28]. With reference to the nomenclature of Ref. [28], we adopt a "kinetic picture" and the Abraham formalism so that the electromagnetic momentum density is $\mathbf{S}/c^2$ [29, 30, 31]. In general, the angular momentum may be decomposed into orbital and spin components [28]. The spin component is origin independent, but for an open system the orbital component (and thereby also the total angular momentum) generally depends on the origin of the coordinate axes [28]. In contrast, for a closed cavity it can be shown that the total light momentum ($\int dV\, \mathbf{S}/c^2$) of a mode always vanishes, and thereby the angular momentum $\mathcal{L}$ is origin independent.

In the supplementary materials [32], it is shown that for a sufficiently large cavity ($d \ll L_i$, $i=1,2$) the $z$-component of the angular momentum is:

$$\frac{\mathcal{L}_z}{A_{tot}} \approx s\frac{2}{c^2 l_P}\left|\int_{\text{cav. perimeter}} dx_\parallel \int dx_\perp \int dz\, S_\parallel\right|, \quad (1)$$

where $A_{tot} = L_1 \times L_2$ is the cross-sectional area, $l_P = 2(L_1 + L_2)$ is the cavity perimeter, and $s = +1$ ($s = -1$) for a mode that circulates in the anti-clockwise (clockwise) direction, respectively. The coordinate $x_\parallel$ is measured along the cavity perimeter and the coordinate $x_\perp$ along the perpendicular direction ($x_\parallel$ and $x_\perp$ are in the *xoy* plane). Furthermore, $S_\parallel$ is the Poynting vector component parallel to the cavity walls. Similar to the theory of photonic crystals [33], the spatially-averaged Poynting vector along the propagation path can be related to the (net) group velocity $v_g$ of the edge-wave and with the energy stored in the cavity $\mathcal{E}$ as follows: $\int_{\text{cav. perimeter}} dx_\parallel \int dx_\perp \int dz\, S_\parallel = v_g \mathcal{E}$ [32]. Hence, the angular momentum can be written as:

$$\frac{\mathcal{L}_z}{A_{tot}} \approx s \frac{2}{c^2 l_P} |v_g| \mathcal{E}. \qquad (2)$$

The correction term that makes the two sides of the equation identical vanishes in the limit $A_{tot} \to \infty$.

Suppose now that the system is in thermodynamic equilibrium with a reservoir at temperature $T$, and that all the light inside the cavity is generated by thermal (or quantum) fluctuations. Crucially, a few recent works [18, 31, 34, 35] have shown that a thermodynamic equilibrium (without any heat sources) is compatible with a circulation of thermal energy in closed orbits. Specifically, for nonreciprocal systems the expectation of the heat current [18, 35] and of the angular momentum [18] can be nontrivial. In the limit of vanishingly small material loss, the angular momentum expectation can be written as $\langle \mathcal{L}_z \rangle = \sum_{\omega_n > 0} \mathcal{E}_{T,\omega_n} \mathcal{L}^{(n)}$ where the summation is over all the positive frequency modes of the cavity. Here, $\mathcal{L}^{(n)}$ is the angular momentum of generic mode normalized to its energy and $\omega_n$ is the oscillation frequency. The formula of $\langle \mathcal{L}_z \rangle$ is consistent with the fluctuation-dissipation theorem (see [18]). The (unilateral) angular momentum spectral density $\mathcal{L}_\omega$ is defined such that $\langle \mathcal{L}_z \rangle = \int_0^\infty d\omega\, \mathcal{L}_\omega$. It has the explicit expression $\mathcal{L}_\omega = \mathcal{E}_{T,\omega} \sum_{\omega_n > 0} \mathcal{L}^{(n)} \delta(\omega - \omega_n)$. The angular momentum expectation may be nontrivial only when the time-reversal symmetry is broken, because otherwise the heat current vanishes ($\left\langle \mathbf{S}/c^2 \right\rangle = 0$) in all points of the cavity [18, 35].

In a band-gap all the cavity modes must be edge-waves. Using Eq. (2) to evaluate the angular momentum ($\mathcal{L}^{(n)}$) of a generic edge-state it is found that:

$$\frac{\mathcal{L}_\omega}{A_{tot}} \approx \mathcal{E}_{T,\omega} \frac{2}{c^2 l_P} \sum_n s_n \left| v_{g,n} \right| \delta(\omega - \omega_n), \qquad \text{(in a band-gap)}. \qquad (3)$$

The summation is over all the (positive-frequency) edge modes in the spectral range of interest. Here, $v_{g,n}$ is the group velocity of the $n$-th mode and $s_n = \pm 1$ depending if the energy flows in the anti-clockwise or clockwise direction. As discussed in the supplementary materials [32], the edge modes can be organized in branches ($m=1,2,\ldots$), and the dispersion of each branch is of the form $\omega_m(k_\parallel)$. The parameter $k_\parallel$ determines the total phase delay ($k_\parallel l_P$) acquired by the wave as it travels one complete loop around the cavity. Even though $k_\parallel$ may be regarded as a continuous variable, only solutions with $k_\parallel l_P = 2\pi n$ with $n$ integer are physical [32]. Thus, for a large cavity perimeter and for a given branch the sum over the edge-waves can be approximated by an integral: $\frac{1}{l_P}\sum_n \to \frac{1}{2\pi}\int dk_\parallel$. This shows that $\frac{\mathcal{L}_\omega}{A_{tot}} \approx \mathcal{E}_{T,\omega} \frac{1}{c^2 \pi} \sum_m \int dk_\parallel s_m \left| v_{g,m} \right| \delta(\omega - \omega_m)$. But since $v_{g,m} = \frac{\partial \omega_m}{\partial k_\parallel}$ [32], we finally conclude that:

$$\frac{\mathcal{L}_\omega}{A_{tot}} \approx -\mathcal{E}_{T,\omega} \frac{1}{c^2 \pi} \mathcal{C}_\omega, \quad \text{with} \quad \mathcal{C}_\omega = -\sum_{\omega_m = \omega} s_m. \qquad (4)$$

The sum is over all the edge-modes for which $\omega_m = \omega$ (each branch may contribute with one or more points). Since $s_m = \pm 1$ the parameter $\mathcal{C}_\omega$ simply counts the difference

-6-

between the number of edge modes associated with an anti-clockwise power flow ($s_m = +1$) and the number of edge modes associated with a clockwise power flow ($s_m = -1$) at frequency $\omega$. In particular, we see that the absolute value of the sum gives the net number of unidirectional modes. This is the first key result of the article. It establishes that in the band-gaps of a photonic insulator cavity, the expectation of the angular momentum spectral-density per unit of area is quantized in units of $\frac{1}{\pi c^2}\mathcal{E}_{T,\omega}$. The quantization is strictly valid in the continuum limit, $A_{tot} \to \infty$ (likewise, in electronic systems the Hall conductivity is quantized only when the sample area approaches infinity). The quantized angular momentum density is determined by the number of unidirectional edge modes ($\mathcal{C}_\omega = 0, \pm 1, \pm 2, ...$) at frequency $\omega$. This property is rather general and does *not* depend on the topological nature of the system. Defects may lead to additional contributions to the angular momentum, but their presence can be safely neglected in state-of-the-art photonic designs. Furthermore, in the supplementary materials it is shown that the power spectral density associated with the energy flow around the cavity walls is also quantized [32].

Let us now focus on topological Chern-type materials [14, 15, 36, 37]. The bulk-edge correspondence principle establishes that the net number of unidirectional edge modes at the interface of two topological materials is coincident with the gap Chern number difference [6, 37]. Thus, supposing that the cavity walls are topologically trivial, (e.g., PEC walls) it follows that $|\mathcal{C}_\omega| = |\mathcal{C}_{gap}|$, where $\mathcal{C}_{gap}$ is the gap Chern number of the photonic insulator. This is the second key result of the article. It implies that the photonic Chern number has a precise physical meaning: it determines the quantized angular momentum spectral density of a cavity in a thermodynamic equilibrium. In particular, we see that the fluctuation-induced angular momentum spectral density will be totally insensitive to any variation of the system parameters (e.g., a change of the biasing magnetic field) that does not close the gap. Note that the Chern number can be unambiguously defined for fully 3D waveguide platforms [38, 39]. Furthermore, a more sophisticated analysis reported elsewhere demonstrates that one has precisely $\mathcal{C}_\omega = \mathcal{C}_{gap}$ (the numerical examples presented in the following confirm this result) [40, 41]. Thus, a positive (negative) gap Chern number implies that the unidirectional edge waves propagate clockwise (anti-clockwise) with respect to the $z$-axis.

To illustrate the application of the developed ideas, we take the photonic insulator as a magnetized-electric-plasma with a gyrotropic permittivity response, $\overline{\overline{\varepsilon}} = \varepsilon_t \mathbf{1}_t + \varepsilon_a \hat{\mathbf{z}} \otimes \hat{\mathbf{z}} + i\varepsilon_g \hat{\mathbf{z}} \times \mathbf{1}$, with $\mathbf{1}_t = \hat{\mathbf{x}} \otimes \hat{\mathbf{x}} + \hat{\mathbf{y}} \otimes \hat{\mathbf{y}}$ and

$$\varepsilon_t = 1 - \frac{\omega_p^2 (1 + i\gamma/\omega)}{(\omega + i\gamma)^2 - \omega_c^2}, \qquad \varepsilon_a = 1 - \frac{\omega_p^2}{\omega(\omega + i\gamma)}, \quad \text{and} \quad \varepsilon_g = \frac{1}{\omega} \frac{\omega_c \omega_p^2}{\omega_c^2 - (\omega + i\gamma)^2}, \quad (5)$$

where $\omega_p$ is the plasma frequency, $\gamma$ is the collision frequency, $\omega_c = -qB_0/m$ is the cyclotron frequency ($\omega_c > 0$ when the magnetic field is directed along +z), $q = -e$ is the electron charge and $m$ is the electron effective mass [42]. The bias magnetic field is $\mathbf{B} = B_0 \hat{\mathbf{z}}$. InSb with a magnetic bias and other narrow-gap semiconductors have a similar gyrotropic response at terahertz frequencies [43, 44].

Figure 2 shows the band structure of a magnetized plasma with $\omega_c = 0.8\omega_p$ (blue lines) for propagation in the *xoy* plane and transverse-magnetic (TM) polarization (the nontrivial field components are $E_x, E_y, H_z$). The medium has two photonic band-gaps. It is known that lossless electromagnetic continua may be topologically classified [36], and in particular a lossless magnetized plasma ($\gamma = 0^+$) has topologically nontrivial phases [18, 45, 46]. The gap Chern numbers $\mathcal{C}_{\text{gap},i}$ indicated in the insets of Fig. 2 are calculated as detailed in Refs. [36, 38]. They include the contributions of all bands below the band-gap, including the negative frequency bands [38] (not shown in Fig. 2). The dispersion of the edge-states supported by a planar interface of the magnetized plasma (region $y>0$) and a PEC material (region $y<0$) is represented by the green lines in Fig. 2. Consistent with the bulk-edge correspondence principle, each band-gap supports a unidirectional edge-wave.

Consider now a cylindrical cavity filled with the magnetized plasma. The cavity lateral walls are PEC. Furthermore, to ease the analytical treatment the cavity cross-section is circular with radius $R$. In the supplementary materials [32] it is shown that the cavity modes have a magnetic field of the form $H_z = H_0 I_{|l|}(\alpha_{ef}\rho)e^{il\varphi}$, with $I_{|l|}$ the modified Bessel function of the 1$^{\text{st}}$ kind, $l = 0, \pm 1, ...$ is the azimuthal quantum number, $\alpha_{ef} = \sqrt{-\varepsilon_{ef}}\,\omega/c$ and $\varepsilon_{ef} = (\varepsilon_t^2 - \varepsilon_g^2)/\varepsilon_t$ is the effective permittivity of the gyrotropic material. The cavity modes satisfy the dispersion equation: $\alpha_{ef}R\dfrac{I'_{|l|}(\alpha_{ef}R)}{I_{|l|}(\alpha_{ef}R)} - l\dfrac{\varepsilon_g}{\varepsilon_t} = 0$. The resonant frequencies for a cavity with radius $R = 10c/\omega_p$ are represented in Fig. 2 as discrete black dots, in the spectral range determined by the band-gaps. As seen, the discrete dots follow closely the dispersion of the edge states associated with a planar interface (green lines), indicating that in the band-gaps the cavity modes are indeed localized near the lateral walls. This property is confirmed by Figs. 3a and 3b, which represent the profiles of two generic cavity modes in the 1$^{st}$ (low-frequency) and 2$^{nd}$ (high-frequency) band-gaps, respectively. The modes of the 1$^{st}$ (2$^{nd}$) gap have a positive (negative) azimuthal quantum number and circulate in the anti-clockwise (clockwise) direction.

The angular momentum spectral density was numerically calculated using the formula $\mathcal{L}_\omega = \mathcal{E}_{T,\omega} \sum_{\omega_n > 0} \mathcal{L}^{(n)} \delta(\omega - \omega_n)$. An explicit expression for the normalized angular momentum of a generic mode ($\mathcal{L}^{(n)}$) is given in Ref. [32]. The $\delta$-function is spread in the interval $(\omega_n + \omega_{n-1})/2 < \omega < (\omega_{n+1} + \omega_n)/2$ as a rectangular pulse with height $2/(\omega_{n+1} - \omega_{n-1})$. Figures 4a and 4b depict the normalized spectral density $\tilde{\mathcal{L}}_\omega \equiv \dfrac{\mathcal{L}_\omega c^2 \pi}{\mathcal{E}_{T,\omega} A_{tot}}$ in the 1$^{st}$ and 2$^{nd}$ band-gaps, respectively, for cavities with radius $R = 10c/\omega_p$ (dot-dashed green lines) and $R = 20c/\omega_p$ (black lines). As seen, even though the cavity radius is only a few free-space wavelengths, the numerical results are consistent with the approximate identity (4): $\tilde{\mathcal{L}}_\omega \approx -\mathcal{C}_\omega = +1$ in the 1$^{st}$ gap and $\tilde{\mathcal{L}}_\omega \approx -\mathcal{C}_\omega = -1$ in the 2$^{nd}$ gap. In $R \to \infty$ limit the approximate identities become exact, and the angular momentum spectral density is exactly quantized and is determined by the gap Chern numbers: $\tilde{\mathcal{L}}_\omega = -\mathcal{C}_{gap,1}$ in the 1$^{st}$ gap and $\tilde{\mathcal{L}}_\omega = -\mathcal{C}_{gap,2}$ in the 2$^{nd}$ gap (dashed blue lines in Fig. 4).

Figure 5 represents the total angular momentum contribution of each band gap, defined as $\langle \mathcal{L}_{z,i} \rangle = \int_{gap\ i} d\omega\, \mathcal{L}_\omega$, as a function of the temperature for $\omega_p/2\pi = 1\,\text{THz}$ and $\omega_p/2\pi = 10\,\text{THz}$. As could be anticipated, the system angular momentum due to the edge modes increases approximately linearly with the temperature because $\mathcal{E}_{T,\omega} \approx k_B T$ for large *T*. The contributions of the band-gaps to the angular momentum have opposite signs. For very low-temperatures the contribution of the 2$^{nd}$ (high-frequency) gap dominates, and the total angular momentum due to edge waves is negative ($\langle \mathcal{L}_{z,1} \rangle + \langle \mathcal{L}_{z,2} \rangle < 0$). In the $T \to 0^+$ limit, the angular momentum is exclusively due to the quantum vacuum fluctuations. In this case, the zero-point energy (due to the gap edge-modes) flows in the clockwise direction around the lateral walls, i.e., in the direction determined by the electron "skipping orbits" (opposite to the bulk cyclotron orbits) [18, 47, 48]. In contrast, for moderately large temperatures the thermal-effects make $\langle \mathcal{L}_{z,1} \rangle + \langle \mathcal{L}_{z,2} \rangle$ positive and the contribution of the 1$^{st}$ (low-frequency) gap dominates. It should be noted that bulk modes may also contribute to the fluctuation induced angular momentum [49, 50], and hence $\langle \mathcal{L}_{z,1} \rangle + \langle \mathcal{L}_{z,2} \rangle$ only gives a partial (band-gap) component of $\langle \mathcal{L}_z \rangle$.

Next, we outline a possible microwave experiment to verify that the fluctuation-induced angular momentum density is nontrivial. Figure 6 depicts the metallic cavity filled with a topological material. A waveguide directional coupler is connected to the cavity through a multihole aperture (similar to a Bethe hole coupler) [51]. In a microwave experiment (e.g., $\omega/2\pi \sim 10\,\text{GHz}$), the topological material can be implemented with a square array of ferrite rods biased with a static magnetic field, similar to Ref. [8]. A small fraction of the thermal energy flowing near the cavity walls may be transferred to either port A or port B of the directional coupler. The directional coupler can be designed in such a manner that when the flow of thermal energy follows an anti-clockwise (clockwise) motion (with respect to the z-axis) most of the energy is coupled to the port B (port A), and the other port is isolated [51].

A measurement must perturb slightly the thermal equilibrium because otherwise the Poynting vector orbits are closed and the thermal flow is impossible to detect (in particular, the zero-point energy part of $\mathcal{E}_{T,\omega}$ cannot be directly measured) [18]. Hence, in an experiment the ports A and B must be held at a temperature $T_0$ different from the cavity temperature $T$. A detailed analysis reported in the supplementary material [32] shows that the difference between the power flows at ports A and B when the thermal equilibrium is perturbed ($T_0 < T$) is given by:

$$\delta p \equiv p_B - p_A \approx k_B (T - T_0) \Delta C_D \frac{\Delta \omega}{2\pi}, \tag{6}$$

where $\Delta \omega$ is the bandwidth of the detecting circuit. The analysis assumes that both ports A and B are terminated by matched loads and that the material dissipation in the directional coupler is negligible due to the weak loss of metals at microwaves. The coefficient $\Delta C_D$ is determined by the difference of the coupling strengths between ports A and B and all the cavity modes; it may be found with an independent calibration process, specifically from the S-parameters of the two-port network [32]. Equation (6) is rather general and applies even when the frequency of interest is outside a band-gap and for any cavity size.

In the following, we concentrate on the case in which the detecting circuit bandwidth lies in a band-gap, so that $\Delta C_D$ is entirely determined by the edge modes. Interestingly, because the power density transported by the edge modes is quantized and is independent of the system area [32], it follows that $\Delta C_D$ will depend very little on the cavity cross-sectional area provided its diameter is a few times larger than the characteristic modal size so that the edge waves attached to opposite walls are well spatially separated. The typical modal size of the edge modes for a topological photonic crystal is on the order of the lattice constant [8]. Furthermore, it will also depend little on the specific location of the directional coupler in the cavity. Thus, a fingerprint of the topological gap is that $\delta p$ must be nearly independent of the cavity cross-section and of the location of the directional coupler. Note that outside a band-gap these properties do not hold.

For a reciprocal cavity its coupling with the two ports is symmetric and hence $\Delta C_D$ vanishes and $\delta p = 0$ [32]. Similarly, for a topologically trivial (but not necessarily reciprocal) gap ($\mathcal{C} = 0$) the number of edge states (if nonzero) circulating in counterclockwise direction is identical to the number of states circulating in the clockwise direction. This indicates that $\Delta C_D$ is negligible when $\mathcal{C} = 0$. Clearly, a significant imbalance between the net energy fluxes at ports A and B is only possible when $\mathcal{C} \neq 0$, and it will provide a clear signature of a nontrivial topology and confirm the circulation of thermal energy in closed orbits.

Ideally the directional coupler should ensure that the power transported by the modes circulating along a given direction is delivered to a single port [32]. For example, if the coupler is designed such that when the edge-modes propagate in an anti-clockwise (clockwise) direction most of the thermal energy is coupled to port B (A) then $\Delta C_D = -\text{sgn}(\mathcal{C})|\Delta C_D|$. Thus, the sign of $\delta p$ is linked to the sign of the Chern number. Furthermore, for an ideal coupler $|\Delta C_D|$ represents the fraction of thermal energy rerouted from the cavity to the coupled port. Standard couplers can be very directive such that the two ports coupling strength differs by four orders of magnitude [51].

At ports A and B the thermal power may be collected by a microwave bolometer (radiometer) [52-53]. The power rerouted from the topological cavity to the coupled port increases the noise temperature of the corresponding bolometer by $\delta T_n \approx T |\Delta C_D|$, where $T$ is the temperature of the cavity [32]. For $|\Delta C_D| \sim 0.1$ and $T = 300K$ the excess of noise temperature on the detector is $\delta T_n \approx 30K$. Thus, the bolometer needs to sufficiently sensitive to detect variations of the noise temperature on the order of some tens of Kelvin. Detectors cooled to a temperature on the order of $T_0 = 4K$ can detect an excess of noise temperature as small as $\sim 1K$ and are of widespread use in radio-astronomy [54, 55]. Therefore, the proposed experiment seems to be within reach using cooled state-of-art bolometers with $T_0 \leq \delta T_n$.

In summary, it was demonstrated that the fluctuation-induced angular momentum in a generic photonic insulator cavity with opaque-type boundaries has a quantized spectral density in the photonic band-gaps. The quantized spectral-density depends on the net number of unidirectional edge-states at the cavity walls. For topological systems, the quantized spectral-density is determined by the photonic Chern number, which thereby has a precise physical meaning as a "quantum" of the light-angular momentum spectral density. For $\mathbb{Z}_2$ topological photonic insulators with the time-reversal symmetry [9] the spin-filtered $\mathcal{L}_\omega / A_{tot}$ is also quantized, but the contributions of the different "spins" cancel out. The nontrivial fluctuation-induced angular momentum may be experimentally verified by coupling the light that circulates around the cavity walls to a directional coupler and by detecting the imbalance between the energy sensed by the different arms of the coupler. We believe that the interpretation of the Chern number as a quantum of the angular momentum spectral density provides a deeper understanding about the intriguing role of topology in photonic systems [41].

**Acknowledgements:** This work is supported in part by Fundação para a Ciência e a Tecnologia grant number PTDC/EEI-TEL/4543/2014 and by Instituto de Telecomunicações under project UID/EEA/50008/2017.

associated with the energy circulating around the cavity walls, and *v)* characterization of the flow of thermal energy from the topological cavity to the arms of the directional coupler.

**Figures:**

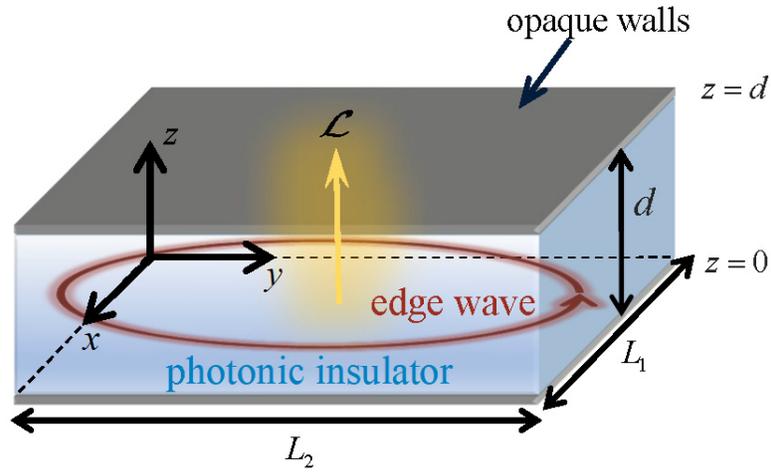

Fig. 1 A closed electromagnetic cavity does not support bulk states in some spectral range ("photonic insulator"). The lateral walls (not shown) may eventually enable the propagation of edge-type waves confined to the boundary. The thermal fluctuation-induced light is generally characterized by a non-trivial angular momentum $\mathcal{L}$, which for a sufficiently large cavity ($A_{tot} = L_1 \times L_2 \to \infty$) has a quantized spectral density in a band-gap.

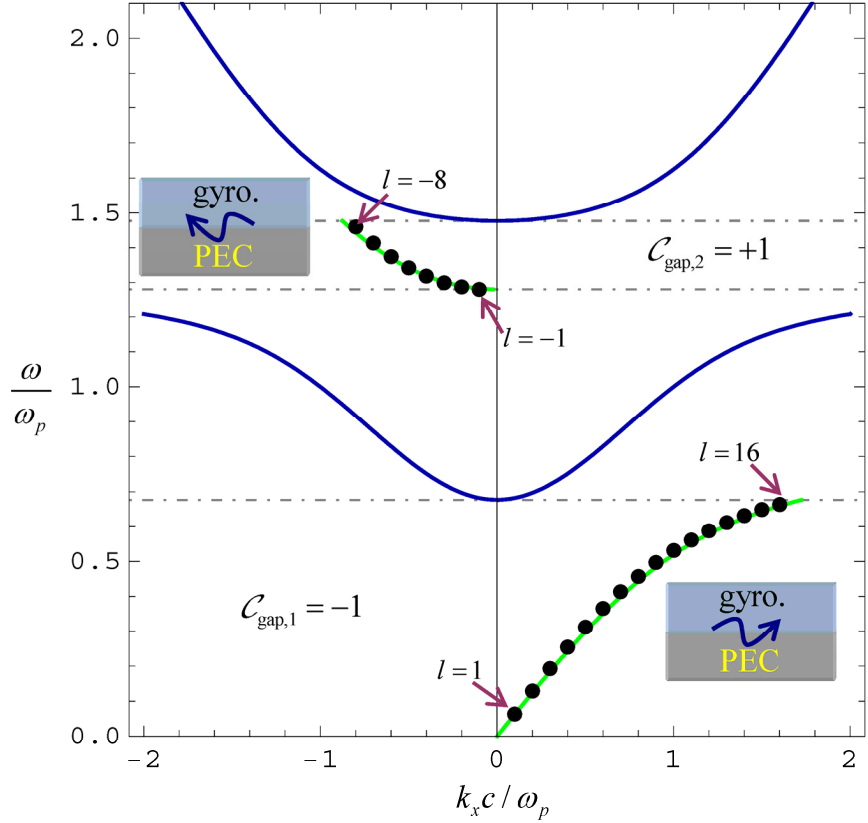

Fig. 2 **Blue lines:** band diagram of the bulk gyrotropic material with $\omega_c = 0.8\omega_p$. The two photonic band-gaps are delimited by the dashed horizontal grid lines. The gap Chern numbers ($\mathcal{C}_{gap,1}$ and $\mathcal{C}_{gap,2}$) are given in the insets. **Green lines:** dispersion of the (gap) edge modes supported by a planar interface ($y=0$) between the gyrotropic material (region $y>0$) and PEC material (region $y<0$). The edge modes propagate along the $x$-axis (see the insets). The edge mode dispersion is shown only in the band-gaps. **Discrete black symbols:** resonant frequencies of a cylindrical cavity ($R = 10c/\omega_p$) with PEC walls filled by the gyrotropic material. The modes are labeled by the azimuthal quantum number $l$.

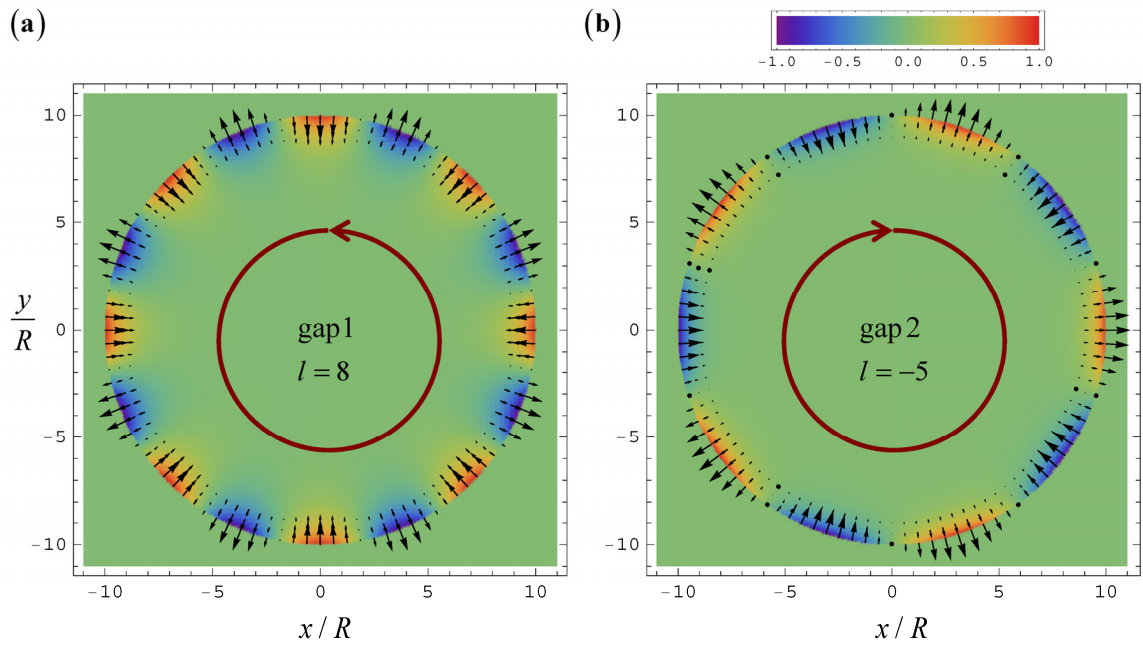

Fig. 3 Density plot of a time snapshot of the magnetic field in the cylindrical cavity ($R = 10c/\omega_p$) for **(a)** the mode $l=8$ in the 1$^{st}$ (low-frequency) gap (propagating in the anti-clockwise direction) and **(b)** the mode $l=-5$ in the 2$^{nd}$ (high-frequency) gap (propagating in the clockwise direction). The arrows represent the electric field lines.

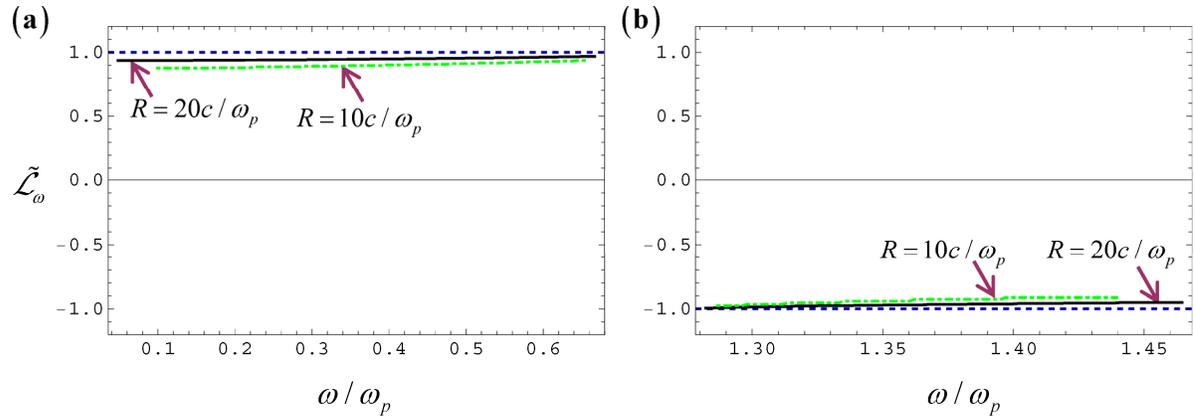

Fig. 4 Normalized angular momentum spectral density in **(a)** the 1st band gap and **(b)** in the 2nd band gap for a cylindrical cavity with radius $R = 10c/\omega_p$ (green dot-dashed line) and $R = 20c/\omega_p$ (black line). As $R \to \infty$, the angular momentum density becomes quantized (blue dashed horizontal line). The cyclotron frequency is $\omega_c = 0.8\omega_p$.

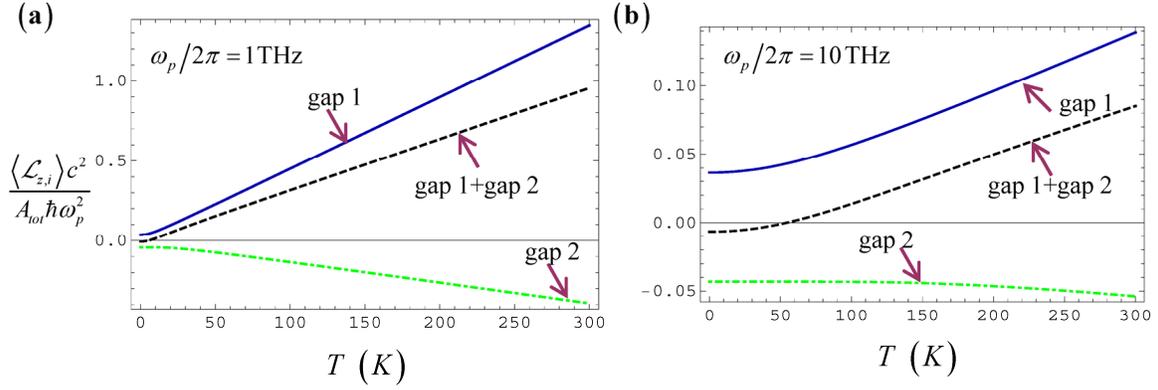

Fig. 5 Partial fluctuation induced angular momentum $\langle \mathcal{L}_{z,i} \rangle$ in a cylindrical cavity with radius $R \gg c/\omega_p$ for **(a)** $\omega_p/2\pi = 1\,\text{THz}$ and **(b)** $\omega_p/2\pi = 10\,\text{THz}$. The cyclotron frequency is $\omega_c = 0.8\omega_p$. Solid blue line: contribution from the 1st (low-frequency) band gap. Dot-dashed green line: contribution from the 2nd (high-frequency) band gap. Dashed black line: combined contribution of the two band gaps. Note that the contribution of the modes outside the spectral region of the band-gaps is not included in the calculation.

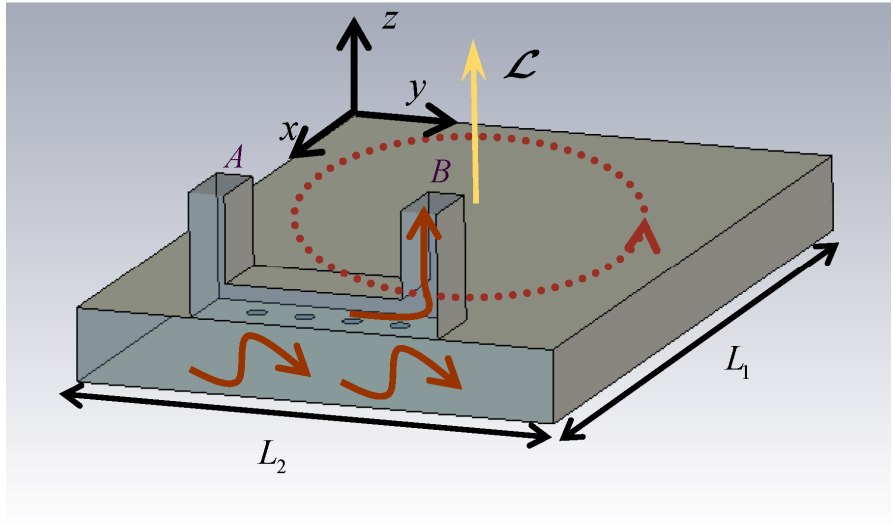

Fig. 6 Setup of a possible experiment at microwaves: A metallic cavity is filled with a topological photonic crystal (not represented in the figure; e.g., a photonic crystal formed by ferrite rods [8]) and is connected to a waveguide coupler through multiple apertures on the top metallic plate. The directional coupler can be designed in such a manner that for an anti-clockwise (clockwise) flow of thermal energy the coupled signal is transmitted mainly to port B (port A). The front wall of the cavity and the front wall of the directional coupler are not shown in order to visualize the interior of the structure.

-24-

# Supplementary material for "Quantized Angular Momentum in Topological Optical Systems"

*Mário G. Silveirinha*[†]

[1] *University of Lisbon–Instituto Superior Técnico and Instituto de Telecomunicações, Avenida Rovisco Pais, 1, 1049-001 Lisboa, Portugal, mario.silveirinha@co.it.pt*

## A. Derivation of the angular momentum of an edge-mode

Let us consider a closed lossless cavity with an edge-mode circulating around the lateral walls (Fig. 1 of the main text). The edge mode is described by the complex-valued electromagnetic field $(\mathbf{E}, \mathbf{H})$ (the time-variation $e^{-i\omega t}$ is implicit). The angular momentum is by definition:

$$\mathcal{L}_z = \frac{1}{c^2} \hat{\mathbf{z}} \cdot \int dV \, \mathbf{r} \times \mathbf{S}, \qquad (A1)$$

where $\mathbf{S} = \mathrm{Re}\{\mathbf{E} \times \mathbf{H}^*\}$ is the Poynting vector of a complex-valued field (the time-averaged Poynting vector of the real-valued field $\mathrm{Re}\{\mathbf{E}e^{-i\omega t}\}$ differs by a $1/2$ factor from $\mathrm{Re}\{\mathbf{E} \times \mathbf{H}^*\}$).

In the general case, the bulk region is a photonic crystal and thus is formed by many identical cells (let us say with dimensions $a_1 \times a_2 \times a_3$). The Poynting vector may vary considerably on the scale of each unit cell. Since the position vector $\mathbf{r}$ varies slowly on this scale, for a large cavity we may write:

$$\mathcal{L}_z \approx \frac{1}{c^2} \hat{\mathbf{z}} \cdot \int dV \, \mathbf{r} \times \mathbf{S}_{\mathrm{av}}, \qquad (A2)$$

---

[†] To whom correspondence should be addressed: E-mail: *mario.silveirinha@co.it.pt*

where $\mathbf{S}_{av}$ should be understood as a spatially averaged Poynting vector with the fluctuations on the scale of the unit cell removed.

We arbitrarily choose the origin of the coordinate axes to be at the center of the cavity. For a large cavity, the edge mode fields are concentrated near the walls and the coupling between different walls is negligible (the contribution of the corners region to the angular momentum is negligible). Hence, $\mathcal{L}_z$ may be regarded as a sum of 4 parcels, with each parcel associated with a specific wall. Let us focus on the wall $x = L_1/2$, which gives the angular momentum contribution $\mathcal{L}_z\big|_{\text{wall}\atop x=L_1/2} = \frac{1}{c^2}\hat{\mathbf{z}} \cdot \int_{\tilde{V}} dV\ \mathbf{r} \times \mathbf{S}_{av}$, where $\tilde{V}$ is some volumetric region nearby the considered wall of the form $L_1/2 - \delta_s \leq x \leq L_1/2$ and $-L_2/2 \leq y \leq L_2/2$. Here, $\delta_s$ is some characteristic penetration depth of the edge mode into the bulk region (the mode has an exponentially decay in the direction perpendicular to the wall). It is implicit that $\delta_s \ll L_i$ ($i=1,2$) so that $\tilde{V}$ does not overlap the edge mode profile associated with other walls. Clearly, $\mathbf{S}_{av}$ must be predominantly oriented along the $\pm \hat{\mathbf{y}}$ direction, because the energy can only flow along directions parallel to the wall. Thus, to leading order it is possible to write

$$\mathcal{L}_z\big|_{\text{wall}\atop x=L_1/2} \approx \frac{1}{c^2}\frac{L_1}{2}\int_{\tilde{V}} dV\ S_{av,y} = \frac{1}{c^2}\frac{L_1}{2}\int_{\tilde{V}} dV\ S_y. \tag{A3}$$

The second identity follows from the fact that the volume integral of a spatially averaged quantity is simply the volume integral of that quantity. One can further write, $\int_{\tilde{V}} dV\ S_y = \int_{-L_2/2}^{L_2/2} dy \left( \int_{L_1/2-\delta_s}^{L_1/2} dx \int dz\ S_y \right)$. The inner integral gives the flux of the Poynting vector through each section $y = const.$, and thereby from the conservation of energy must be independent of $y$. This shows that:

$$\mathcal{L}_z\Big|_{\substack{\text{wall}\\x=L_1/2}} \approx \frac{1}{c^2}\frac{L_1 L_2}{2}\int\limits_{L_1/2-\delta_s}^{L_1/2} dx \int dz\, S_y. \quad (A4)$$

Proceeding in the same way for the other 3 walls, we obtain similar formulas. Generically, the contribution of a given wall is of the form $\frac{1}{c^2}\frac{A_{tot}}{2}\int dx_\perp \int dz\, S_\parallel$, where $x_\perp$ is the coordinate perpendicular to the interface and $S_\parallel$ is the Poynting vector component parallel to the interface (positive when the Poynting vector is oriented in the anti-clockwise direction, with respect to $z$). For a large cavity, the integral $\int dx_\perp \int dz\, S_\parallel$ is independent of the considered section-cut (around the cavity perimeter) due to the conservation of energy. Hence,

$$\begin{aligned}\frac{\mathcal{L}_z}{A_{tot}} &\approx \frac{2}{c^2}\int dx_\perp \int dz\, S_\parallel \\ &= s\frac{2}{c^2 l_P}\left|\int\limits_{\text{cav. perimeter}} dx_\parallel \int dx_\perp \int dz\, S_\parallel\right|.\end{aligned} \quad (A5)$$

where $s = \pm 1$ for modes that circulate in the anti-clockwise (clockwise) direction and $l_P = 2(L_1 + L_2)$ is the cavity perimeter.

### B. Edge-mode branches and group velocity

In order to characterize the edge modes supported by a closed cavity, we introduce an auxiliary mathematical problem whose geometry is coincident with the original one, except that the fields are allowed to be discontinuous near some rectangular cut ($x = 0$, $-L_2/2 < y < -L_2/2 + w$, and $0 < z < d$) near the inferior lateral wall (see Fig.

S1). We will refer to this region simply as the "discontinuity plane". It is implicit that the oscillation frequency is in a band-gap of the bulk region. The width $w$ should be large enough so that it exceeds the penetration depth of the edge modes into the bulk region ($w > \delta_s$) and at the same time $w << L_2/2$ (thus, the cavity needs to be sufficiently large).

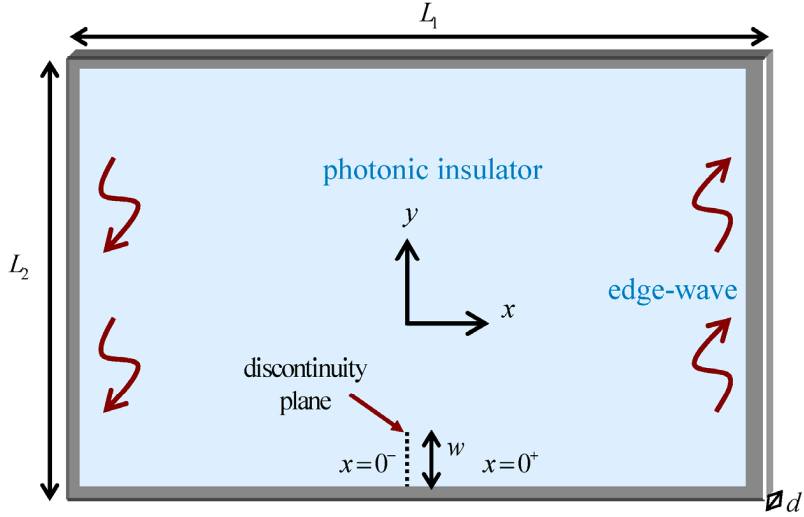

Fig. S1 Geometry of the auxiliary problem: a discontinuity plane (represented by the dashed line) is inserted near one of the lateral walls. Bloch-type boundary conditions are enforced at the discontinuity plane [Eq. (B1)]. The edge modes propagate around the cavity perimeter and cannot penetrate into the bulk region, which is a photonic insulator.

We are interested in the eigen-solutions of the Maxwell's equations such that the fields at the "+" and "−" sides of the discontinuity plane are linked by Bloch-type boundary conditions:

$$\mathbf{E}_{\tan}\big|_{x=0^-} = \mathbf{E}_{\tan}\big|_{x=0^+} e^{ik_\| l_P}, \qquad \mathbf{H}_{\tan}\big|_{x=0^-} = \mathbf{H}_{\tan}\big|_{x=0^+} e^{ik_\| l_P}. \tag{B1}$$

The subscript "tan" refers to the field components tangential to the discontinuity plane. The parameter $k_\|$ determines the phase delay ($k_\| l_P$) acquired by the wave as it goes around the lateral walls of the cavity with perimeter $l_P = 2(L_1 + L_2)$. Evidently, solutions with $k_\| l_P = 2\pi n$ and $n$ integer are also solutions of the original problem with no discontinuity plane. However, here it is convenient to admit that $k_\|$ can take any real value. Furthermore, by analytic continuation, we can also consider eigen-solutions with $k_\|$ complex, $k_\| = k' + ik''$, which are evidently associated with some complex-valued eigen-frequencies $\omega = \omega' + i\omega''$.

Let us then consider a generic family of natural modes $\left(\mathbf{E}_{k_\|}, \mathbf{H}_{k_\|}\right)$ of the Maxwell's equations in the cavity that satisfy the Bloch-type boundary conditions (B1) with dispersion $\omega = \omega(k_\|)$.

For a lossless cavity, the conservation of energy implies that $\nabla \cdot \mathbf{S} + \partial_t W = 0$, with $\mathbf{S}$ the Poynting vector and $W$ the instantaneous stored energy density. For a dispersive lossless medium and complex-valued fields the Poynting vector is defined as $\mathbf{S} = \mathrm{Re}\{\mathbf{E} \times \mathbf{H}^*\}$, whereas the electromagnetic energy density may be written as $W = \frac{1}{2}\mathbf{Q}^* \cdot \mathbf{M}_g \cdot \mathbf{Q}$, with $\mathbf{Q}$ the state-vector of the system and $\mathbf{M}_g$ a generalized material matrix [R1, R2]. Integrating the formula $\nabla \cdot \mathbf{S} + \partial_t W = 0$ over the cavity volume, it is found that:

$$\frac{d}{dt}\int dV\, W(\mathbf{r},t) = \hat{\mathbf{x}} \cdot \int_{\substack{\text{discont.}\\ \text{plane}}} dS \left[\mathbf{S}(\mathbf{r},t)\big|_{x=0^+} - \mathbf{S}(\mathbf{r},t)\big|_{x=0^-}\right]. \tag{B2}$$

For complex-valued fields with a time variation $e^{-i\omega t} = e^{-i\omega' t} e^{\omega'' t}$ both $\mathbf{S}$ and $W$ vary in time as $e^{2\omega'' t}$. Furthermore, for modes that satisfy Eq. (B1) with a complex-valued $k_\| = k' + ik''$ it is evident that $\mathbf{S}(\mathbf{r},t)\big|_{x=0^-} = e^{-2k''l_P} \mathbf{S}(\mathbf{r},t)\big|_{x=0^+}$. Hence, for a generic eigenmode it is possible to write:

$$2\omega'' \int dV\, W_{k'+ik''}(\mathbf{r},t) = \left(1 - e^{-2k''l_P}\right) \hat{\mathbf{x}} \cdot \int_{\substack{\text{discont.}\\ \text{plane}}} dS\, \mathbf{S}_{k'+ik''}(\mathbf{r},t)\big|_{x=0^+}, \tag{B3}$$

with $\omega' + i\omega'' = \omega(k_\|)\big|_{k_\| = k' + ik''}$. Next, we take the limit $k'' \to 0$. Since the eigenmodes satisfy the dispersion equation $\omega = \omega(k_\|)$, by doing a Taylor expansion it is seen that $\omega'' = \dfrac{\partial \omega}{\partial k_\|}(k') k''$. Hence, modes with a real-valued $k_\|$ satisfy:

$$v_g \int dV\, W_{k'}(\mathbf{r}) = l_P\, \hat{\mathbf{x}} \cdot \int_{\substack{\text{discont.}\\ \text{plane}}} dS\, \mathbf{S}_{k'}(\mathbf{r})\big|_{x=0^+}. \tag{B4}$$

where $v_g = \dfrac{\partial \omega}{\partial k_\|}(k')$ may be understood as the (net) group-velocity of the edge-mode. Note that for a real-valued oscillation frequency both $\mathbf{S}$ and $W$ are independent of time. Moreover, the conservation of energy implies that the integral $\hat{\mathbf{x}} \cdot \int_{\substack{\text{discont.}\\ \text{plane}}} dS\, \mathbf{S}_{k'}\big|_{x=0^+} = \int_{\substack{\text{discont.}\\ \text{plane}}} dx_\perp dz\, S_\|\big|_{x=0^+}$ is independent of the section-cut around perimeter of the cavity (see Sect. A). Thus, a generic edge-mode of the original cavity satisfies (dropping all the irrelevant subscripts):

$$\int_{\text{cav. perimeter}} dx_\| \int dx_\perp \int dz\, S_\| = v_g \int_{\text{cavity}} dV\, W = v_g \mathcal{E}, \tag{B5}$$

where $\mathcal{E}$ is the stored energy. The notations $\|$ and $\perp$ are used with the same meaning as in Sect. A. The above formula generalizes the result $\int dV\, \mathbf{S} = \mathbf{v}_g \int dV\, W$, satisfied by generic Bloch waves in photonic crystals [R3, R4].

## C. Modes of a cylindrical cavity filled with a gyrotropic material

We consider a cylindrical cavity with radius $R$ filled with a gyrotropic material with (relative) permittivity tensor $\overline{\varepsilon} = \varepsilon_t \mathbf{1}_t + \varepsilon_a \hat{\mathbf{z}} \otimes \hat{\mathbf{z}} + i\varepsilon_g \hat{\mathbf{z}} \times \mathbf{1}$, with $\mathbf{1}_t = \hat{\mathbf{x}} \otimes \hat{\mathbf{x}} + \hat{\mathbf{y}} \otimes \hat{\mathbf{y}}$, and a trivial permeability. The cavity lateral wall is a perfect electric conducting (PEC) surface. The fields are assumed to be TM-polarized with $\mathbf{H} = H_z \hat{\mathbf{z}}$ and $\mathbf{E} = E_x \hat{\mathbf{x}} + E_y \hat{\mathbf{y}}$. It is implicit that the fields are independent of the $z$-coordinate ($0 < z < d$) so that the problem is effectively two-dimensional. In the bulk region (the interior of the cavity) the magnetic field satisfies [R5]:

$$\left( \nabla^2 + \frac{\omega^2}{c^2} \varepsilon_{ef} \right) H_z = 0, \qquad \text{with} \quad \varepsilon_{ef} = \frac{\varepsilon_t^2 - \varepsilon_g^2}{\varepsilon_t}. \tag{C1}$$

Hence, adopting a system of cylindrical coordinates $(\rho, \varphi)$ it is clear that a generic natural mode has a magnetic field of the form (apart from an arbitrary normalization factor):

$$\eta_0 H_z = I_{|l|}(\alpha_{ef} \rho) e^{il\varphi}, \tag{C2}$$

where $\eta_0$ is the free-space impedance, $I_{|l|}$ the modified Bessel function of the 1$^{\text{st}}$ kind, $l = 0, \pm 1, \ldots$ is the azimuthal quantum number and $\alpha_{ef} = \sqrt{-\varepsilon_{ef}}\, \omega/c$. The electric field can be found using $-i\omega\varepsilon_0 \mathbf{E} = \overline{\varepsilon}^{-1} \cdot \nabla \times \mathbf{H}$ with $\overline{\varepsilon}^{-1} = \frac{1}{\varepsilon_{ef}}\left( \mathbf{1}_t - i\frac{\varepsilon_g}{\varepsilon_t} \hat{\mathbf{z}} \times \mathbf{1} \right) + \frac{1}{\varepsilon_a} \hat{\mathbf{z}} \otimes \hat{\mathbf{z}}$. Since $\nabla \times \mathbf{H} = \frac{1}{\rho} \partial_\varphi H_z \hat{\boldsymbol{\rho}} - \partial_\rho H_z \hat{\boldsymbol{\varphi}}$ it follows that:

$$\mathbf{E} = \frac{1}{i\omega\varepsilon_0 \varepsilon_{ef}} \left( -\frac{1}{\rho} \partial_\varphi H_z + \frac{i\varepsilon_g}{\varepsilon_t} \partial_\rho H_z \right) \hat{\boldsymbol{\rho}} + \frac{1}{i\omega\varepsilon_0 \varepsilon_{ef}} \left( \partial_\rho H_z + \frac{i\varepsilon_g}{\varepsilon_t} \frac{1}{\rho} \partial_\varphi H_z \right) \hat{\boldsymbol{\varphi}}. \tag{C3}$$

Using Eq. (C2), we obtain the explicit formula:

$$\mathbf{E} = \frac{1}{i\varepsilon_{ef}\omega/c}\left(-il\frac{1}{\rho}I_{|l|}(\alpha_{ef}\rho) + \frac{i\varepsilon_g}{\varepsilon_t}\alpha_{ef}I'_{|l|}(\alpha_{ef}\rho)\right)\hat{\boldsymbol{\rho}}$$
$$+ \frac{1}{i\varepsilon_{ef}\omega/c}\left(\alpha_{ef}I'_{|l|}(\alpha_{ef}\rho) - l\frac{\varepsilon_g}{\varepsilon_t}\frac{1}{\rho}I_{|l|}(\alpha_{ef}\rho)\right)\hat{\boldsymbol{\varphi}} \quad (C4)$$

Imposing the PEC boundary condition ($E_\varphi = 0$) at $\rho = R$, one finds the dispersion equation for the natural modes:

$$\alpha_{ef}R\frac{I'_{|l|}(\alpha_{ef}R)}{I_{|l|}(\alpha_{ef}R)} - l\frac{\varepsilon_g}{\varepsilon_t} = 0. \quad (C5)$$

The Poynting vector ($\mathbf{S} = \mathrm{Re}\{\mathbf{E}\times\mathbf{H}^*\}$) of a (complex-valued) mode has the azimuthal component $S_\varphi = -\mathrm{Re}\{E_\rho H_z^*\}$. The angular momentum of the mode is $\mathcal{L}_z = \frac{1}{c^2}\int dV \rho S_\varphi$, which from Eqs. (C2) and (C4) can be written explicitly as ($d$ is the height of the cavity along $z$):

$$\frac{\mathcal{L}_z}{d} = \frac{2\pi}{\eta_0}\frac{1}{c^2}\frac{1}{\varepsilon_{ef}\omega/c}\int_0^R d\rho\,\rho\,\mathrm{Re}\left\{I^*_{|l|}(\alpha_{ef}\rho)\left(lI_{|l|}(\alpha_{ef}\rho) - \frac{\varepsilon_g}{\varepsilon_t}\alpha_{ef}\rho I'_{|l|}(\alpha_{ef}\rho)\right)\right\}. \quad (C6)$$

The energy density of the complex–valued field is (note that for time-averaged real-valued fields one needs to multiply the right-hand side by an additional $1/2$ factor) [R1, R6]:

$$W = \frac{1}{2}\varepsilon_0\left(\mathbf{E}^*\cdot\partial_\omega(\omega\bar{\varepsilon})\cdot\mathbf{E} + |\eta_0 H_z|^2\right). \quad (C7)$$

Thus, after some straightforward calculations the stored energy, $\mathcal{E} = \int dV W$, can be written as:

$$\frac{\mathcal{E}}{d} = 2\pi \int_0^R d\rho \frac{1}{2} \varepsilon_0 \rho \left( \left( \partial_\omega (\omega \varepsilon_t) \right) \left( |E_\rho|^2 + |E_\varphi|^2 \right) + 2 \partial_\omega (\omega \varepsilon_g) \operatorname{Re}\{iE_\rho E_\varphi^*\} + |\eta_0 H_z|^2 \right), \quad \text{(C8)}$$

where $\partial_\omega = \partial/\partial\omega$. Combining Eqs. (C6) and (C8) one finds that the angular momentum of a cavity mode normalized to its energy is given by:

$$\frac{\mathcal{L}_z}{\mathcal{E}} = \frac{1}{\omega} \frac{\frac{1}{\varepsilon_{ef}} \int_0^R d\rho \, \rho \operatorname{Re}\left\{ I_{|l|}^*(\alpha_{ef}\rho) \left( l I_{|l|}(\alpha_{ef}\rho) - \frac{\varepsilon_g}{\varepsilon_t} \alpha_{ef} \rho I_{|l|}'(\alpha_{ef}\rho) \right) \right\}}{\int_0^R d\rho \frac{1}{2} \rho \left( \left( \partial_\omega (\omega \varepsilon_t) \right) \left( |E_\rho|^2 + |E_\varphi|^2 \right) + 2 \partial_\omega (\omega \varepsilon_g) \operatorname{Re}\{iE_\rho E_\varphi^*\} + |\eta_0 H_z|^2 \right)}, \quad \text{(C9)}$$

with $E_\rho, E_\varphi, H_z$ defined as in Eqs. (C2) and (C4). The right-hand side of the above equation determines the parameter $\mathcal{L}^{(n)}$ used in the main text.

### D. Power spectral density circulating around the cavity walls

Let $p_\omega$ be the unilateral power spectral density associated with the energy transported around the cavity walls ($P = \int_0^\infty d\omega \, p_\omega$). From Eq. (B5), the power transported by a given edge mode is:

$$P = \frac{1}{l_P} v_g \mathcal{E}, \quad \text{(D1)}$$

where $l_P$ is the cavity perimeter. Thus, the power spectral density in a band-gap is given by

$$p_\omega = \mathcal{E}_{T,\omega} \frac{1}{l_P} \sum_n v_{g,n} \delta(\omega - \omega_n), \quad \text{(D2)}$$

with the summation over all the edge modes. For a sufficiently large cavity we can use $\frac{1}{l_P} \sum_n \to \frac{1}{2\pi} \int dk_\parallel$ (see the main text). Taking into account the contributions of all edge-mode branches and the link between the gap Chern number and the net number of unidirectional edge modes one obtains:

$$p_\omega \approx -\mathcal{C}\frac{1}{2\pi}\mathcal{E}_{T,\omega}. \tag{D3}$$

Thus, it follows that $p_\omega$ is also quantized in units of $\frac{1}{2\pi}\mathcal{E}_{T,\omega}$. Note that the sign of $p_\omega$ determines the direction (anti-clockwise vs. clockwise) along which the thermal energy flows.

In simple terms, in the band gaps the cavity is analogous to a one-dimensional circular transmission line. Indeed, the fluctuation-induced power density transported by a standard transmission line mode is precisely $p_\omega^{\text{per mode}} = \pm\frac{1}{2\pi}\mathcal{E}_{T,\omega}$ [R7]. In a nonreciprocal line, the net number of unidirectional modes can be nonzero, leading to a nontrivial $p_\omega$. Note that from the definition of angular momentum it follows immediately that for a circular transmission line $\mathcal{L}_\omega = \frac{2A_{tot}}{c^2}p_\omega$.

### E. Power rerouted from the cavity to the directional coupler

In the following, we characterize the net power rerouted from the topological cavity (at temperature $T$) to the arms of the directional coupler (Fig. 6 of the main text). It is supposed that the material loss in the directional coupler is negligible and that ports A and B are terminated with matched loads (microwave bolometers) cooled to a temperature $T_0 < T$. For simplicity, in the following we refer to ports A and B as ports 1 and 2, respectively.

The thermal energy radiated by port 1 ($p_{rad,1}$) is $p_{rad,1} = \left(1 - |s_{11}|^2\right) \varepsilon_{T_0,\omega}^{ther} \frac{\Delta\omega}{2\pi}$, where $\Delta\omega$ is the relevant bandwidth. Here, $s_{ij}$ are the scattering parameters of the two-port microwave network. Furthermore, $\varepsilon_{T,\omega}^{ther} = \frac{\hbar\omega}{2}\left[\coth\left(\frac{\hbar\omega}{2k_B T}\right) - 1\right] = \frac{\hbar\omega}{e^{\hbar\omega/k_B T} - 1} \approx k_B T$ is the mean thermal energy of a harmonic oscillator at temperature $T$. Note that $\varepsilon_{T,\omega}^{ther} \frac{\Delta\omega}{2\pi}$ is the thermal noise power delivered by a circuit with temperature $T$ to a matched load [R8].

On the other hand, the thermal energy captured by port 1 ($p_{abs,1}$) is $p_{abs,1} = |s_{12}|^2 \varepsilon_{T_0,\omega}^{ther} \frac{\Delta\omega}{2\pi} + \left(1 - |s_{11}|^2 - |s_{12}|^2\right) \varepsilon_{T,\omega}^{ther} \frac{\Delta\omega}{2\pi}$. Note that $|s_{12}|^2 \varepsilon_{T_0,\omega}^{ther} \frac{\Delta\omega}{2\pi}$ is the power collected at port 1 due to the thermal radiation from port 2. The leading coefficient of the second term of $p_{abs,1}$ is found by noting that at thermal equilibrium ($T = T_0$) one needs to have $p_{abs,1} = p_{rad,1}$. Using $\varepsilon_{T,\omega}^{ther} \approx k_B T$ it follows that the power collected by ports $i=1,2$ is given by:

$$p_{abs,1} = |s_{12}|^2 k_B T_0 \frac{\Delta\omega}{2\pi} + C_{3\to 1} k_B T \frac{\Delta\omega}{2\pi}. \tag{E1a}$$

$$p_{abs,2} = |s_{21}|^2 k_B T_0 \frac{\Delta\omega}{2\pi} + C_{3\to 2} k_B T \frac{\Delta\omega}{2\pi}. \tag{E1b}$$

where $C_{3\to 1} = 1 - |s_{11}|^2 - |s_{12}|^2$ and $C_{3\to 2} = 1 - |s_{22}|^2 - |s_{21}|^2$ are the coefficients that determine the coupling between the cavity (index 3) and the ports 1 and 2, respectively. The net power flow $p_i = p_{abs,i} - p_{rad,i}$ at port $i$ can be written as:

$$p_i = C_{3\to i} k_B (T - T_0) \frac{\Delta\omega}{2\pi}. \tag{E2}$$

Thus, $\delta p \equiv p_2 - p_1$ is given by $\delta p = (C_{3\to 2} - C_{3\to 1}) k_B (T - T_0) \frac{\Delta \omega}{2\pi}$. This yields the result of the main text with $\Delta C_D = C_{3\to 2} - C_{3\to 1}$.

The directional coupler may be designed such that one of the coefficients $C_{3\to i}$ is negligible in a band-gap. For example, if the directional coupler ensures that when the topological modes rotate in the counter-clockwise direction ($\mathcal{C} < 0$) most of the energy is coupled to port B, then we can write $\delta p = -\text{sgn}(\mathcal{C}) k_B (T - T_0) |\Delta C_D| \frac{\Delta \omega}{2\pi}$. Note that the coefficients $C_{3\to i}$ are determined by the scattering parameters of the microwave network and hence can be easily determined with a vector network analyzer. In the reciprocal case $s_{21} = s_{12}$ and by symmetry $s_{11} = s_{22}$ so that $\delta p = 0$.

The equivalent noise temperature ($T_{n,i}$) at port $i$ is defined such that $p_{abs,i} = k_B T_{n,i} \frac{\Delta \omega}{2\pi}$ [R8]. From Eq. (E1) it is possible to write $T_{n,1} = |s_{12}|^2 T_0 + C_{3\to 1} T$ and $T_{n,2} = |s_{21}|^2 T_0 + C_{3\to 2} T$. If the coupling with the topological cavity is removed the noise temperature is $T_n = T_0$ (note that the ports 1 and 2 are terminated with matched loads). Hence, the excess of thermal noise in port 2 due to the coupling with the cavity is $\delta T_{n,2} = -(1 - |s_{21}|^2) T_0 + C_{3\to 2} T$. For a weak cavity coupling, and a detector with a sufficiently low temperature the first term is negligible ($(1 - |s_{21}|^2) T_0 \ll T_0$) provided the second term is comparable to or larger than $T_0$. In these conditions, the excess of noise temperature in the coupled port is roughly $\delta T_n \approx |\Delta C_D| T$.